    \documentclass[12pt,psf,epsf]{article}
\usepackage[centertags]{amsmath}
\usepackage{amssymb}
\usepackage{graphicx}
\usepackage{epsfig}
\usepackage[english]{babel}
\usepackage{array}
\usepackage{amsthm}
\usepackage{latexsym}
\usepackage[mathcal]{euscript}
\pdfoutput=1
\usepackage{epsfig}

 \usepackage{jheppub}
 \usepackage{hyperref}

\usepackage{subfigure}
\usepackage{psfrag}

\usepackage{epsfig}
\usepackage[latin1]{inputenc}
\usepackage{float}
\usepackage{graphicx}
\usepackage{cancel}
\usepackage{mathrsfs}
\usepackage{amssymb}
\usepackage{amsfonts}
\usepackage{amsmath}
\usepackage{slashed}
\usepackage{graphicx}
\usepackage{bm}
\usepackage{color}
\newcommand{\be}{\begin{equation}}
\newcommand{\ee}{\end{equation}}
\newcommand{\bea}{\begin{eqnarray}}
\newcommand{\eea}{\end{eqnarray}}
\newcommand{\bwt}{\begin{widetext}}
\newcommand{\ewt}{\end{widetext}}

\newcommand{\bi}{\begin{itemize}}
\newcommand{\ei}{\end{itemize}}

\def\Tr{\mathop{\rm Tr}}

\newcommand{\rar}{\rightarrow}

\usepackage{setspace}

\begin{document}

\title {Anisotropic destruction of the Fermi surface in inhomogeneous holographic lattices}

\author[a,d]{Askar Iliasov,}
\author[a,b]{Andrey A. Bagrov,}
\author[a]{Mikhail I. Katsnelson,}
\author[c]{\hbox{Alexander Krikun}}
\affiliation[a]{Institute for Molecules and Materials, Radboud University, Heyendaalseweg 135, 6525AJ Nijmegen, \mbox{The Netherlands}}
\affiliation[b]{Department of  Physics  and  Astronomy,  Uppsala  University, Box 516,  SE-75120  Uppsala,  Sweden}
\affiliation[c]{Instituut-Lorentz  for  Theoretical  Physics, $\Delta$-ITP,  Leiden  University, Niels  Bohrweg 2, 2333CA Leiden, The Netherlands}
\affiliation[d]{Space Research Institute of the Russian Academy of Science, Moscow, 117997, Russia}

\emailAdd{a.iliasov@science.ru.nl}
\emailAdd{andrey.bagrov@physics.uu.se}
\emailAdd{m.katsnelson@science.ru.nl}
\emailAdd{krikun@lorentz.leidenuniv.nl}

\abstract{
We analyze fermionic response of strongly correlated holographic matter in presence of inhomogeneous periodically modulated potential mimicking the crystal lattice. The modulation is sourced by a scalar operator that explicitly breaks the translational symmetry in one direction. 
We compute the fermion spectral function and show that 
it either exhibits a well defined Fermi surface with umklapp gaps opening on the Brillouin zone boundary at small lattice wave vector, or,  
when the wave vector is large, the Fermi surface is anisotropically deformed and the quasiparticles get significantly broadened in the direction of translation symmetry breaking. 

Making use of the ability of our model to smoothly extrapolate to the homogeneous Q-lattice like setup, we show that this novel effect is not due to the periodic modulation of the potential and Umklapp physics, but rather due to the anisotropic features of the holographic horizon. That means it encodes novel physics of strongly correlated critical systems which may be relevant for phenomenology of exotic states of electron matter.
}

\maketitle

\newpage

\section{Introduction}
Phenomenological treatment of quasiparticle decoherence is hindered by the lack of theoretical tools. While the Landau Fermi liquid is conceptually unable to treat it, the robust non-Fermi liquid theory is yet missing. However, there is a handy tool capable to access the behavior of  
strongly coupled quantum systems at least qualitatively 
-- the holographic duality \cite{Zaanen:2015oix, keimer2015quantum}.  The holographic approach to model building
resides on the correspondence between  
operators in the strongly coupled quantum field theory and 
auxiliary classical gravitational system involving a black hole in a space-time with one extra dimension. The fundamental objects which one can study in this way are correlation function of quantum operators, i.e. the Green's functions. This construction does not assume the existence of stable quasiparticles in the boundary field theory. However, they may emerge as poles of the holographically evaluated Green's function. This emergent nature of quasiparticle excitations in 
holography makes it convenient for studying transitions between conventional Fermi-liquid and ``exotic''  non-Fermi-liquid states of quantum matter.

The observable which can be directly accessed with experimental probes, and in particular with the angle-resolved photoemission spectroscopy (ARPES) is momentum-resolved spectral function, related to the fermionic Green's function
\begin{equation}
\rho (\omega, k) = \mbox{Im} \Tr \langle \Psi^\dag \Psi \rangle.    
\end{equation}
It has been shown in the pioneering works \cite{Cubrovic:2009ye,Cubrovic:2011xm,Liu:2009dm,Faulkner:2009wj,Faulkner:2010tq} that this object can be evaluated holgraphycally by introducing 
fermionic field $\psi$ in the gravitational bulk, which is dual to 
fermionic operator $\Psi$ on the boundary. The classical solutions to the corresponding Dirac equation encode the fermionic Green's function. The striking observation of \cite{Faulkner:2009wj,Faulkner:2011tm,Cubrovic:2009ye} is that, depending on the parameters of the model, the Green's function may or may not have poles -- the coherent quasiparticle excitations. 
In real strongly correlated systems like superconducting cuprates the quasiparticle decoherence manifests itself in anisotropic deformations of the Fermi surface leading to nodal-antinodal dichotomy \cite{fu2006dichotomy}, or to complete destruction of quasiparticles and emergence of Fermi arcs \cite{kanigel2006evolution}. A bit less known but also interesting example is provided by electron states near the Fermi energy in half-metallic ferromagnets which have quasiparticle character for one spin projection and non-quasiparticle for the other one \cite{katsnelson2008}. This poses a question whether one can build a holographic model with the coexistence of quasiparticle and non-quasiparticle behavior, and especially, keeping in mind possible application to the cuprates, the model where the lifetime of quasiparticle excitations would be significantly anisotropic.

In real quantum systems the anisotropy is caused by the crystal lattice which breaks the rotation symmetry down to a discrete subgroup. It is natural to introduce the crystal lattice in the holographic model as well, in order to obtain the anisotropic fermionic spectral function. Holographic models with periodic external potentials have been introduced in \cite{Horowitz:2012ky, Horowitz:2012gs} and further studied in \cite{Donos:2014yya, Rangamani:2015hka}. The fermionic response function in these setups has been first addressed in \cite{Liu:2012tr,Ling:2013aya} where it was shown that holography gives non-pathological results. The structure of the Brillouin zone (BZ) created by the periodic potential was reproduced, and gaps opened due to Umklapp scattering on the BZ boundary. Recently, a more subtle result has been reported in \cite{cremonini2018holographic,Cremonini:2019fzz}, where it was claimed that in the periodic lattices generated by intertwined  
order parameters the Fermi surface gets anisotropically destroyed. 

On the other hand, the translation and rotation symmetry breaking can be realized in a class of holographic models which do not possess the periodic potentials. These are so-called holographic homogeneous lattices:  the linear axion model \cite{Andrade:2013gsa}, the Q-lattice model \cite{Donos:2013eha}, and the Bianchy VII helical model \cite{Donos:2012js}. These theoretical constructions are somewhat artificial, but they provide a valuable tool for studying the effects of the translation symmmetry breaking without extra complications 
caused by spatial inhomogeneity of the solutions, so that their technical treatment is greatly simplified. The fermionic response of these models has been studied in \cite{Ling:2014bda} (Q-lattice) and \cite{Bagrov:2016cnr} (helical model). In both cases it was observed that the qusiparticle excitations of the Fermi surface acquire significant width due to the translation symmetry breaking and the pattern of decoherence is anisotropic in the anisotropic backgrounds. The phenomenological significance of these results however has been questioned since the relevance of the homogeneous holographic lattices for real crystal systems has never been completely understood.

Apart from conventional crystals, the effects of periodic potential are relevant in the new class of systems called Van der Waals heterostructures consistent of relatively weakly coupled layers of two-dimensional materials \cite{geim2013}, such as graphene on hexagonal boron nitride (hBN) \cite{dean2010,xue2011,woods2014} or twisted bilayer graphene \cite{cao1,cao2}. In the latter case correlation effects play obviously an important role resulting in metal-insulator transition \cite{cao1} and, possibly, in unconventional superconductivity \cite{cao2}. Moreover, there are some experimental evidences that this system is close to the ``Planck regime'' characteristic of holographic strange metals \cite{cao3}. The advantage of these new crystals is the tunability of the periodic potential which can be modified in a broad range by a simple rotation of one crystal with respect to another \cite{woods2014}. 

The action of the periodic modulation on electronic structure of graphene in Van der Waals heterostructures includes both changes of on-site electrostatic potential and modulation of metrics via the change of hopping parameters \cite{wallbank2013,slotman2015}. The models of homogeneous lattices take into account only the second effect. Despite it does not give the full quantitative description of the problem it is always useful to have an exact solution for a model which takes into account an important part of the total physical picture.

The aim of the current study is to clarify the relation between the anisotropic Fermi surface destruction observed in the periodic lattice model of \cite{cremonini2018holographic,Cremonini:2019fzz} and the similar effect found earlier in homogeneous lattices \cite{Ling:2014bda,Bagrov:2016cnr}. We use the specific model where the periodic potential is introduced via modulation of the extra scalar operator \cite{Horowitz:2012ky}. As has been shown in \cite{vanacore2014minding}, by introducing an extra scalar operator with the opposite phase of modulation this model can be continuously deformed into homogeneous Q-lattice. Therefore it provides a way to disentangle the effects of the Umklapp scattering due to periodicity of the potential, which would disappear when the model is driven to the homogeneous state, and the effects caused by anisotropic breaking of translations, which will remain unaffected. As we will show, the phenomenon observed in the periodic lattices of \cite{cremonini2018holographic,Cremonini:2019fzz} has precisely the same nature as the one seen in the Q-lattice of \cite{Ling:2014bda} thus providing a deeper understanding of which aspects of holographic models are relevant for phenomenology of strongly correlated condensed matter systems.

The paper is organized as follows. In Secs.\,\ref{sec:setup} and \ref{sec:fermions}, we introduce the holographic model with two periodic scalar fields which interpolates between 
homogeneous and inhomogeneous settings and outline the procedure we use to evaluate the fermionic spectral function. We present our main results in Sec.\,\ref{sec:Spectral} and provide a discussion in Sec.\,\ref{sec:Disc}. App. \ref{app:numerical} contains details of the employed numerical scheme. In App.\,\ref{app:extra_results} and \ref{app:q1}, some additional numerical results are presented.

This work complements a similar study \cite{BalmLattice}, where the holographic fermionic spectral function is computed for the periodic ionic lattice. However our setup differs from \cite{BalmLattice} by 
the freedom to extrapolate to the homogeneous lattice, and 
the physical effect which is in focus of \cite{BalmLattice} is quite different from what we are studying in this work.

\section{\label{sec:setup}Holographic non-homogeneous scalar lattice}
We consider the holographic model with two scalar fields that allows to study both periodic and homogeneous settings \cite{vanacore2014minding}:
\begin{gather}\label{eq:action}
	S=\frac{1}{16\pi G}\int d^4 x \sqrt{-g}\left[ R+\frac{6}{L^2}-\partial_\mu \phi \partial^\mu \phi-\partial_\mu \chi \partial^\mu \chi-\frac{1}{2}F^{\mu\nu}F_{\mu\nu}-2V(\phi, \chi)\right],\\
	\label{equ:potentials}
	V(\phi, \chi)=-(\phi^2+\chi^2)/L^2
	\end{gather}
Here $R$ is the Ricci scalar, the cosmological constant is equal to $\Lambda = -3/L^2$ and $L$ is the curvature radius of the resulting 3+1-dimensional AdS space-time. In what follows we will set the gravitational constant to $16\pi G = 2$. The Abelian gauge field is described by the field strength tensor $F=dA$ and the two scalar fields are neutral. The choice of the scalar potential corresponds to $m_{\phi,\chi}^2=-2/L^2$.
The action (\ref{eq:action}) leads to the equations of motion:
		\begin{gather}\label{eq:ein_eq}
	R_{\mu\nu}+\frac{3}{L^2}g_{\mu\nu}-\Big(\partial_\mu \phi \partial_\nu \phi+\partial_\mu \chi \partial_\nu \chi - V(\phi,\chi)g_{\mu\nu}\Big)-\Big(F_{\mu\lambda}F_\nu^\lambda-\frac{g_{\mu\nu}}{4}F_{\lambda\rho}F^{\lambda\rho}\Big)=0,\\
	\nabla_\mu F^\mu_\nu=\frac{1}{\sqrt{-g}}\partial_\mu \left(\sqrt{-g}F^{\mu\sigma}\right)g_{\sigma\nu}=0, \nonumber \\
	\frac{1}{\sqrt{-g}}\partial_\mu\left(\sqrt{-g}\partial^\mu \phi\right) - \frac{\partial V(\phi,\chi)}{\partial\phi}=0,\nonumber\\
	\frac{1}{\sqrt{-g}}\partial_\mu\left(\sqrt{-g}\partial^\mu \chi\right) - \frac{\partial V(\phi,\chi)}{\partial\chi}=0\nonumber
	\end{gather}
Note that with our choice of the scalar potential \eqref{equ:potentials} asymptotic behavior of the scalar fields near the  
conformal boundary $z\rar0$ is described by the two branches
\begin{align}
\phi(x,z)\Big|_{z\rar 0} &= z \phi^{(1)}(x) + z^2 \phi^{(2)}(x) + \dots, \\
\chi(x,z)\Big|_{z\rar 0} &= z \chi^{(1)}(x) + z^2 \chi^{(2)}(x) + \dots,
\end{align}
where 
coefficients $\phi^{(1)}(x)$, $\chi^{(1)}(x)$ in the leading branches, in the direct quantization scheme, are associated to the sources 
of the corresponding dual scalar operators in the boundary theory. If we choose these sources to be spatially dependent, this will introduce explicit translational symmetry breaking in the model.

We will set them to
\begin{align}
\label{eq:interpolation}
\phi^{(1)}(x) = \cos(\theta) V_0\cos(k_0 x), \\
\notag
\chi^{(1)}(x) = \sin(\theta) V_0\sin(k_0 x). 
\end{align}
The phase $\theta$ will allow us to interpolate between the homogeneous and the inhomogeneous setups. Indeed, if we take $\theta=0$, then the $\chi$ field is trivial, 
and we are left with a single periodic scalar source on the boundary with amplitude $V_0$, -- the inhomogeneous setup discussed in \cite{Horowitz:2012ky,Rangamani:2015hka}. However, when we turn $\theta = \pi/4$, the two scalar fields can be thought of as the real and the imaginary components of a single complex scalar field $\Phi = \phi + \chi$ with boundary source 
\begin{equation}
\Phi \big|_{z \rar 0} = z V_0 e^{i k_0 x}.   
\end{equation}
This is equivalent to the Q-lattice setup \cite{Donos:2013eha}, where the $x$-dependence drops out from the equations of motion since they are insensitive to the phase of the complex scalar. Therefore in this case we will get the homogeneous model where the Brillouin zone in not defined, and the corresponding Umklapp scattering is absent. 

Following \cite{Horowitz:2012ky}, we employ the Ansatz for the metric tensor:
	\begin{gather}
	ds^2=\frac{L^2}{z^2} \left[-(1-z)P(z)Q_{tt}dt^2+\frac{Q_{zz}dz^2}{(1-z)P(z)}+Q_{xx}(dx+z^2 Q_{xz}dz)^2+Q_{yy}dy^2 \right].\label{eq:Q-ansatz} \\
	P(z)=1+z+z^2-\frac{\mu^2 z^3}{2}
	\end{gather}
	Here the black hole horizon is located at $z_h=1$, and the Hawking temperature is 
	\begin{equation}
	\label{equ:temperature}
	T=\frac{P(1)}{4\pi L}=\frac{6-\mu^2}{8\pi L},
	\end{equation}
	where $\mu$ is the boundary theory chemical potential related to the charge of the RN black hole.
	The corresponding equations that emerge if \eqref{eq:Q-ansatz} is substituted into \eqref{eq:ein_eq} are too complicated to be written out explicitly. 
	
	Since Einstein equations are not elliptic, the boundary value problem is ill-posed. Therefore in order to impose the boundary conditions on the horizon and the conformal boundary one has to bring first the equations into elliptic form by adding the so called DeTurck tensor that fixes the gauge dynamically \cite{Headrick:2009pv,Wiseman:2011by,Adam:2011dn}:
	\begin{equation}
	G^{H}_{\mu\nu}=G_{\mu\nu}-\nabla_{(\mu}\xi_{\nu)},
	\end{equation}
	where $G_{\mu\nu}$ is the Einstein tensor, $\xi^{\mu}=g^{\lambda\rho}[\Gamma^{\mu}_{\lambda\rho}(g)-\overline{\Gamma}^{\mu}_{\lambda\rho}(\overline{g})]$ and
	$\overline{\Gamma}^{\mu}_{\lambda\rho}(\overline{g})$ is the Levi-Civita connection for some reference metric $\overline{g}$ that should have the same asymptotics and horizon structure as $g$. In our calculations, we use the Reissner-Nordstr\"om (RN) metric for this purpose, hence boundary conditions on the conformal boundary look rather simple:
	\begin{gather}\label{eq:confbond}
	Q_{tt}(0,x)=Q_{xx}(0,x)=Q_{yy}(0,x)=Q_{zz}(0,x)=1\\\nonumber
	Q_{xz}(0,x)=0, \,\, A_t(0,x)=\mu.
	\end{gather}
	
	On the other hand, in order to obtain the boundary conditions on the horizon one should extract the leading and the subleading asymptotics of the equations of motion as $z\rightarrow 1$ and impose that 
	the equations are satisfied, assuming that the field profiles are regular. This gives a set of generalized Robin-like boundary conditions relating the $z$-derivatives of the functions to their values at the horizon together with one algebraic relation $Q_{zz}(1,z) = Q_{tt}(1,z)$.
	
	Once the DeTurck transformation is performed and the boundary conditions are fixed, equations (\ref{eq:ein_eq}) need to be solved numerically. To represent functions and their derivatives on the grid we use the pseudospectral method. Namely, to define finite difference scheme along $z$ axis we employ the method of Chebyshev polynomials, while finite differences along $x$ axis are defined by means of the Fourier expansion. Further, the equations are solved with the Newton-Raphson method. A description of this approach can be found in \citep{Wiseman:2011by}, and we present details of our numerical scheme in Appendix\,\ref{app:numerical}.
	
\section{\label{sec:fermions}Fermionic response and Dirac equation}
	We are interested in fermionic response of the holographic system described above. To derive it, we solve the Dirac equation \cite{Cubrovic:2009ye,Liu:2009dm}
	on the background obtained by solving \eqref{eq:ein_eq}:
	\begin{equation}\label{eq:Dirac}
	\Gamma^{\underline{a}}e^{\mu}_{\underline{a}} \left(\partial_{\mu}+\frac{1}{4}\omega_{\underline{ab}\mu}\Gamma^{\underline{ab}}-iqA_{\mu}\right)\zeta-m\zeta=0.
	\end{equation}
	Here $e^{\mu}_{\underline{a}}$ is the vielbein, $\omega_{\underline{ab}\mu}$ is the spin connection, $\Gamma^{\underline{ab}}=\frac{1}{2}[\Gamma^{\underline{a}},\Gamma^{\underline{b}}]$ is the commutator of gamma matrices, $A_{\mu}$ is the bulk electromagnetic potential, $q$ and $m$ are the charge and the mass of the bulk fermion correspondingly. We adopt the following choice of gamma matrices:
		\begin{gather*}\label{eq:Gmatrices}
	\Gamma^{\underline{t}}=\begin{pmatrix}i\sigma_1 & 0 \\ 0& i\sigma_1\end{pmatrix}, \,\,\,\,
		\Gamma^{\underline{x}}=\begin{pmatrix}-\sigma_2 & 0 \\ 0& -\sigma_2\end{pmatrix},
			\Gamma^{\underline{y}}=\begin{pmatrix}0 & \sigma_2 \\ \sigma_2& 0\end{pmatrix}, \,\,\,\,
				\Gamma^{\underline{z}}=\begin{pmatrix}-\sigma_3 & 0 \\ 0& -\sigma_3\end{pmatrix},
	\end{gather*}
		where $\sigma_{i}$ are Pauli matrices. 
		
	Note that equation (\ref{eq:Dirac}) written in the original Poincare coordinates, with $z_b=0$ and $z_h=1$, exhibits a non-analytic behavior near horizon. It can be shown that, if one expands the solution for the spinor $\zeta$ in powers of $(1-z)$, it will contain half-integer powers. To overcome this, we make a coordinate transformation $z\rightarrow (1-r^2)$. Since we have the background solution only in a numerical form, we perform this transformation pointwise (and thus deform the grid) and then interpolate the results to obtain the values of the bulk functions on the Chebyshev grid suitable for the analysis of the fermionic problem. Non-zero elements of vielbein in the new coordinates are 
\\
	\begin{gather*}\label{eq:verbein}
	e^t_{\underline{t}}=\frac{1-r^2}{r\sqrt{P(r)Q_{tt}}}, \,\,\,\, e^x_{\underline{x}}=\frac{1-r^2}{\sqrt{Q_{xx}}}, \,\,\,\,e^y_{\underline{y}}=\frac{1-r^2}{\sqrt{Q_{yy}}},\\
	e^z_{\underline{x}}=-r(1-r^2)^3 Q_{xz}\sqrt{\frac{P(r)}{Q_{zz}}}, \,\,\,\
	e^z_{\underline{z}}=-\frac{(1-r^2)}{2}\sqrt{\frac{P(r)}{Q_{zz}}},
	\end{gather*}
	where $P(r) \equiv P(z(r))$, 
	and the near-horizon expansion of $\zeta$ reads (note that the horizon is located at $r=0$ and the boundary is at $r=1$):
	\begin{equation}
	\zeta=r^{\pm2i\omega/4\pi LT}(\zeta_0(x)+\zeta_1(x)r+\zeta_2(x)r^2+...).
	\end{equation}
	Out of the two possibilities we choose the factor $r^{-2i\omega/4\pi LT}$, which corresponds to the in-falling boundary condition. 
	
	Function $r^{-2i\omega/4\pi LT}$ is non-analytic, and there are other non-analytic multipliers coming from the near-boundary asymptotics. Therefore we should further redefine $\zeta$ in order to get rid of them:
	\begin{gather*}\label{eq:redefine}
	\zeta=e^{-i(\omega t-x k_x-y k_y)}r^{-2i\omega/4\pi LT}(1-r^2)^{3/2}\Big(P(r)Q_{tt}(r)Q_{xx}(r)Q_{yy}(r)\Big)^{-\frac{1}{4}}\begin{pmatrix}\Psi_1(x,r) \\ \Psi_2(x,r) \end{pmatrix}
	\end{gather*}
	where $\Psi_\alpha$, with $\alpha=1,2$, are two-component spinors, and we added explicit dependence on frequency $\omega$, momentum $k_y$ and quasi-momentum $k_x$.
		Dirac equations for $\Psi$ then read:
		\begin{align}
	 \label{eq:Dirac_main} 
		&\left(\partial_r+  2r\big( 1-r^2 \big)^2 Q_{xz} \partial_x  +\Pi_1 - \frac{2mr}{1-r^2}\frac{Q_{zz}}{P(r)}\sigma_3  -i\Pi_2\sigma_2 \right) \Psi_1 \\
		\notag
		&\qquad + \left( -2ir\frac{P(r)Q_{zz}}{Q_{xx}}(\partial_x +\Pi_3)\sigma_1 \right) \Psi_1
		+ \left(\frac{ik_y (1-r^2)}{\sqrt{Q_{yy}}}\sigma_2-2r\frac{Q_{zz}}{P(r)Q_{yy}}\sigma_3 \right)  \Psi_2=0 \\
		\notag
		& \left(\partial_r+ 2r\big( 1-r^2 \big)^2 Q_{xz} \partial_x + \Pi_1 - \frac{2mr}{1-r^2}\frac{Q_{zz}}{P(r)}\sigma_3 - i\Pi_2 \sigma_2 \right) \Psi_2 \\
		\notag
		&\qquad + \left(+ 2ir\frac{P(r)Q_{zz}}{Q_{xx}}(\partial_x+\Pi_3)\sigma_1 \right) \Psi_2
		+\left(\frac{ik_y (1-r^2)}{\sqrt{Q_{yy}}}\sigma_2-2r\frac{Q_{zz}}{P(r)Q_{yy}}\sigma_3 \right)\Psi_1=0,
	\end{align}
		where
	\begin{align*}
	\Pi_1&=r(1-r^2)^2[2ik_xQ_{xz}+\partial_xQ_{xz}]-\frac{2i\omega}{4\pi LT r}, \\
	\Pi_2&=\frac{2r}{P(r)}\sqrt{\frac{Q_{zz}}{Q_{tt}}}(\omega+q\mu r^2 A_{tt}), \qquad  
	\Pi_3=ik_x-\frac{\partial_xQ_{xx}}{4Q_{xx}}+\frac{\partial_xQ_{zz}}{4Q_{zz}}.
	\end{align*}
Near the conformal boundary ($r\rar 1$) $\Psi$ can be expanded as:
	\begin{gather*}
	\label{equ:boundary_femrion_asymptotes}
	\Psi_\alpha (x) =a_\alpha(x) (1-r)^{-mL}\begin{pmatrix} 0 \\ 1 \end{pmatrix}+b_\alpha(x) (1-r)^{mL}\begin{pmatrix} 1 \\ 0 \end{pmatrix}+...
	\end{gather*}
In the direct quantization scheme, the coefficient in front of the leading branch of the solution, $a_\alpha(x)$, is associated with the source of the dual fermionic operator and the subleading one, $b_\alpha(x)$, 
-- with the response \cite{Liu:2009dm}. For simplicity, in what follows we will be dealing with massless bulk fermion, $m=0$.

In the linear response approximation, the fermionic Green's function allows one to evaluate the response given a particular source:
	\begin{gather}
	\begin{pmatrix} b_1(x) \\ b_2(x) \end{pmatrix}= \int dx' \ G^{R}(\omega,k_x,k_y|x,x') \begin{pmatrix} a_1(x') \\ a_2(x') \end{pmatrix}. \label{eq:ferm_green}
	\end{gather}

Note that in presence of a periodic lattice, the source and the response profiles keep the periodic dependence on the $x$-coordinate even after factoring out the Bloch momentum, therefore the expression above involves a convolution in position space, and $G^R$ is a function of both $x$ and $x'$, with parameters $\omega, k_y$ and $k_x$.  

Taking into account that all the $x$-dependent functions under consideration are periodic with the lattice wave-vector 
$k_0$, it is convenient to expand the sources and the responses in Fourier series
\begin{align}
 a_\alpha (x) &= \sum_{l=-\infty}^{\infty} a_\alpha^l e^{i k_0 x l},\\
 b_\alpha (x) &= \sum_{m=-\infty}^{\infty} b_\alpha^m e^{i k_0 x m}.
\end{align}

In this representation the Green's function $G^R(\omega,k_x,k_y|x,x')$ turns into a matrix in space of the Fourier modes and we get 
\begin{gather}
\begin{pmatrix} b^m_1(x) \\ b^m_2(x) \end{pmatrix}= \sum_{l} \ \left(G^{R} \right)^m_{l} \begin{pmatrix} a^l_1(x') \\ a^l_2(x') \end{pmatrix}.
\end{gather}

In practice, the most phenomenologically relevant part of the $\left(G^{R} \right)^m_{l}$ matrix is the $00$ component, since it characterizes the relation between the plain wave part of the source and the plain wave part of the response on the fermionic perturbation. The reason is that in  anlge-resolved photoemission spectroscopy (ARPES) experiments, which we are interested in, both the excitation -- the infalling photon -- and the measured object -- the photo-electron -- are plain waves. Their overlap with the higher crystal modes can be neglected and therefore the central object of our study is the $\left(G^{R} \right)^0_{0}$ element of the Green's function in Fourier representation (see also the more detailed discussion in \cite{BalmLattice}).

Using \eqref{eq:ferm_green}, it is straightforward to evaluate this 
object. We impose the homogeneous boundary conditions for the fermionic sources $a_1(x) = 1$, or $a_2(x)=1$, depending on the spin component under consideration, and after solving the Dirac equation numerically we extract the homogeneous component of the response term 
from the field profile $b_\alpha(x)$ according  to \eqref{equ:boundary_femrion_asymptotes}. Recall that the element $\left(G^{R} \right)^0_{0}$ is by itself a matrix in the spin representation, therefore 
by taking the source vector $(a_1(x),\,a_2(x))$ to be either $(1,\,0)$ or $(0,\,1)$, we obtain the corresponding values of $b_1$ and $b_2$ that give us columns of the boundary fermionic Green's function. 

In order to compute the retarded correlator, we impose the infalling wave boundary conditions at the horizon \cite{Son:2002sd}. After substituting these modes into the near-horizon expansion of the Dirac equations \eqref{eq:Dirac_main}, we obtain two algebraic and two generalized boundary conditions at horizon. Together with the two Dirichlet boundary conditions for the fermionic sources at $r=1$ and two extra relations coming from the expansion of the equations near the AdS boundary, we acquire a set of 4+4 boundary conditions, which is just enough to formulate the numerical boundary value problem.\footnote{Note however, that the Dirac equations are the first-order differential equations, and setting eight boundary conditions for four equations makes the system overdetermined. Nonetheless, since half of our boundary conditions are obtained from the expansion of the equations of motion themselves, they are always consistent with a solution and therefore never introduce any extra constraints.}

\section{Spectral properties of holographic fermions}\label{sec:Spectral}
Having defined all the components required to conduct the analysis, we can obtain the fermionic spectral function 
\begin{equation}
    A(\omega,k)= \mbox{Im} \Tr G^R(\omega,k)
\end{equation}
by solving \eqref{eq:Dirac_main} on the background given by \eqref{eq:Q-ansatz}.
In this section, we present our numerical results. We set the bulk fermion mass and charge to
\begin{equation}
m=0, \qquad q=1.5,
\end{equation}
(for the results at smaller charges see App.\,\ref{app:extra_results}). To avoid the singularities arising at zero temperature and frequency, in our calculations we use 
\begin{equation}
\omega=10^{-5} \mu, \qquad T=0.01 \mu.
\end{equation}
\begin{figure}
		\centering
		\includegraphics[width=0.7\textwidth]{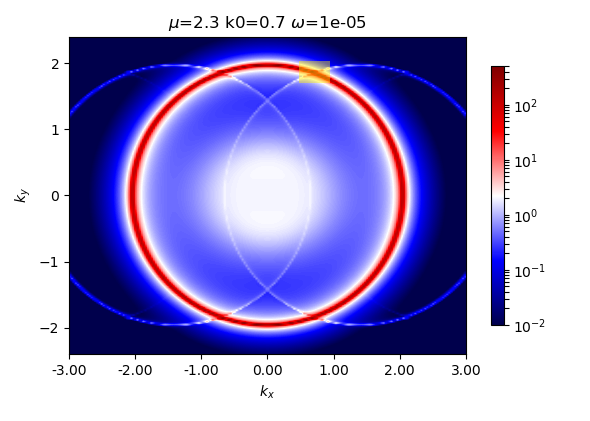}\\
		\includegraphics[width=0.7\textwidth]{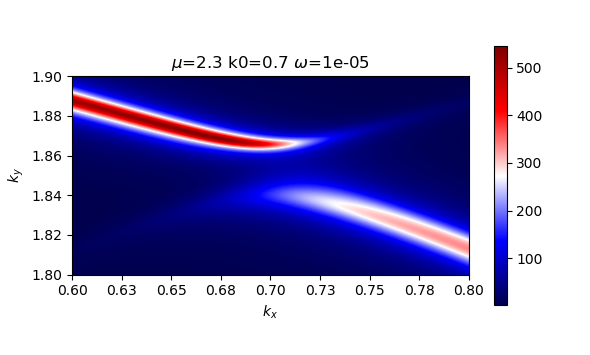}
		\caption{
		\label{fig:shadow_surfaces}
		Upper plot: fermion spectral function for the periodic scalar lattice background at $k_0=0.7 \mu$, $V_0=7$, $q=1.5$, logarithmic scale. The secondary Fermi surfaces arising due to the periodic potential are visible. Yellow rectangle around intersection of the primary and the secondary Fermi surfaces is zoomed and showed in the bottom plot (linear scale). Opening of the Umklapp gap is evident.}
		\end{figure}
We start by considering a background with a single periodic scalar source, $\theta = 0, V_0 = 7$ in \eqref{eq:interpolation} and choose the lattice wave vector $k_0 = 0.7 \mu$, which is smaller than the Fermi momentum $k_F$, so that the Brillouin zone is smaller than the Fermi surface. 
Periodic lattice generally leads to appearance of the suppressed secondary Fermi surfaces in the neighboring Brillouin zones, as one can see on the upper plot of Fig.\,\ref{fig:shadow_surfaces}. When the Fermi surface is larger than the first Brillouin zone, a band gap opens due to Umklapp scattering at the instersection point with a secondary Fermi surface (bottom plot of Fig.\,\ref{fig:shadow_surfaces}). This effect is well-known in the context of conventional condensed matter theory as electronic topological transition, or Lifshitz transition \cite{lifshitz1973,vonsovsky1989}, and holographic calculations reproduce it well, as it has been shown already in \cite{Liu:2012tr,Ling:2013aya,cremonini2018holographic,BalmLattice}. Since in our model the potential is sourced by a neutral scalar, and the coupling between the bulk fermion and the lattice occurs indirectly via modulations of the metric, the amplitude of the secondary surfaces, as well as the size of the band gap are considerably smaller than in the case of charged scalar \cite{cremonini2018holographic}. Nonetheless, the fact that the gap is present in our setting provides a nontrivial test of validity of our treatment. On the other hand, since the effect is weak, this conventional single-particle phenomenon does not obstruct our 
analysis of novel effects caused by the lattice on strongly correlated holographic matter.

\begin{figure}
\center
\begin{tabular}{cc}
\begin{minipage}{0.2 \linewidth}
Periodic \\  lattice \\ $\theta = 0$
\end{minipage}
 &
 \begin{minipage}{0.5 \linewidth}
 \center
\includegraphics[width=1\linewidth]{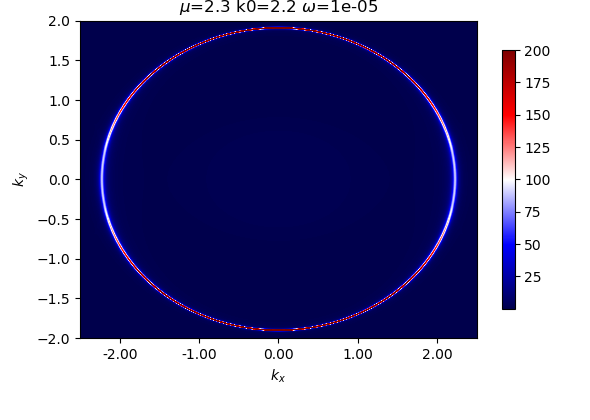}
\end{minipage}
\\
\begin{minipage}{0.2 \linewidth}
Intermediate \\ regime \\ $\theta=\pi/8$
\end{minipage}
&
\begin{minipage}{0.5 \linewidth}
\center
\includegraphics[width=1\linewidth]{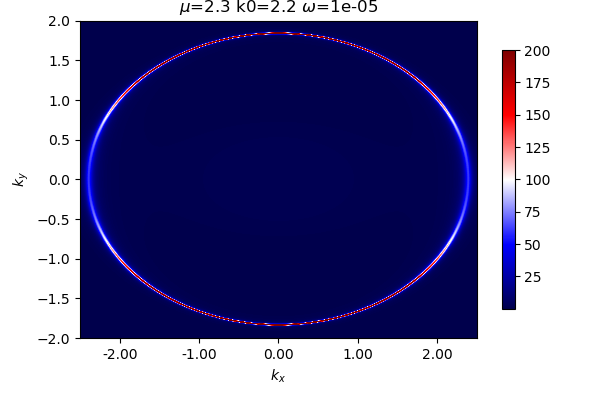}  
\end{minipage}
\\
\begin{minipage}{0.2 \linewidth}
Q-lattice \\ $\theta=\pi/4$
\end{minipage}
&
\begin{minipage}{0.5 \linewidth}
\center
\includegraphics[width=1\linewidth]{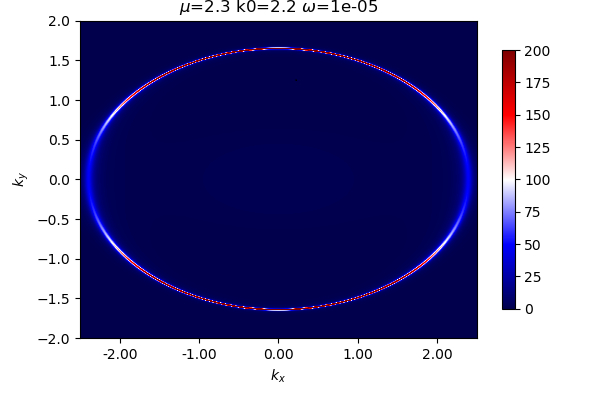}
\end{minipage}
\end{tabular}
\caption{\textbf{Interpolation between the periodic scalar potential and the Q-lattice} $k_0=2.2$, $V_0 = 7$, $q=1.5$. The interpolation is realized by tuning the $\theta$ parameter \eqref{eq:interpolation}, which controls the amplitude of the extra off-phase scalar potential. The umklapp surfaces (out of the scope) completely disappear in the homogeneous Q-lattice setup, while the anisotropic suppression of the Fermi surface is qualitatively unchanged.
\label{fig:anisotropy_interpolation}
}
\end{figure}

When we increase the lattice wave vector, the Brillouin zone gets bigger, the Fermi surfaces do not intersect anymore, and there is no Umklapp gap. 
In general, based on the single-particle intuition, one would expect the effect of periodic lattice to decrease. However, for large enough wave vector $k_0 = 2.2 \mu$, we observe anisotropic stretching of the Fermi surface and suppression of the spectral density in the direction, where translational symmetry is broken, Fig.\,\ref{fig:anisotropy_interpolation} (upper part).
This phenomenon is the central object of our study. First, we note that it is quite similar to the partial suppression of the spectral weight on the Fermi surface observed earlier in \cite{Cremonini:2019fzz}. Indeed, our model is very similar to \cite{Cremonini:2019fzz} since it also has an inhomogeneous holographic lattice and a scalar field. In contrast to \cite{Cremonini:2019fzz}, however, in our case the Brillouin zone boundary is far away, the single-particle Umklapp effects do not interfere with holographic many-body physics, and the phenomenon of anisotropic decoherence is seen much more clearly.

Given that the Brillouin zone boundary doesn't seem to play any discernible role here, it is interesting to figure out whether this effect has anything to do with anisotropic suppression of the Fermi surface observed earlier in the homogeneous Q-lattice model \cite{Ling:2014bda} and, to some extent, in the Bianchy VII helix model \cite{Bagrov:2016cnr}. Indeed, our setting is amenable to smooth interpolation between periodic potential and Q-lattices by tuning parameter $\theta$ in \eqref{eq:action}-\eqref{eq:interpolation}, and we can readily compute how the Fermi surface evolves upon transition from $\theta=0$ (periodic potential) to $\theta=\pi/4$ (Q-lattice).

\begin{figure}
				\includegraphics[width=0.5\textwidth]{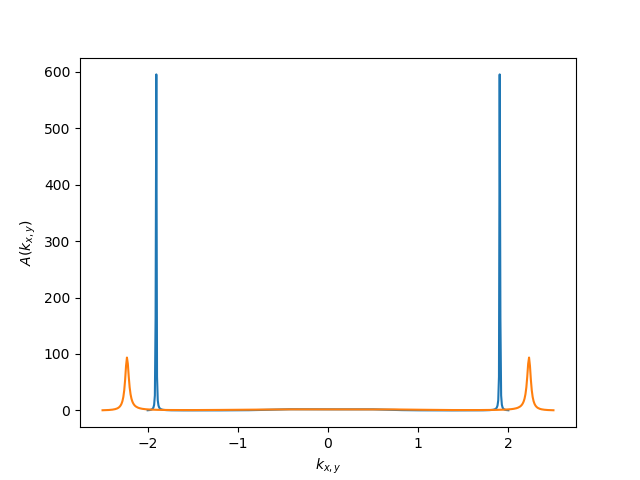}
				\includegraphics[width=0.5\textwidth]{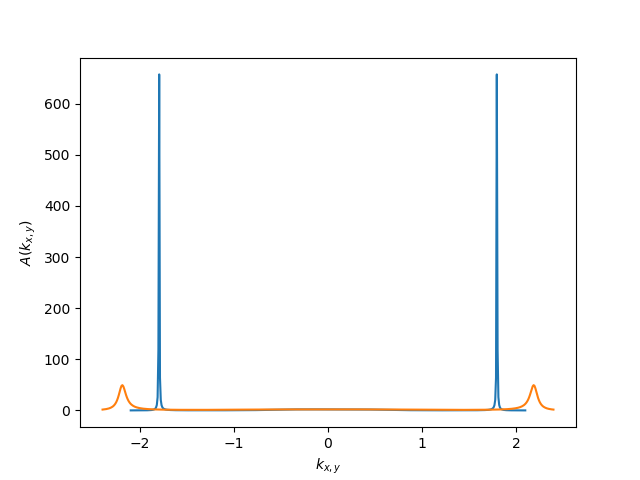}
\caption{\textbf{1d sections of the spectral function} for periodic potential (left) and for Q-lattice (right) at $k_0=2.2$, $V_0=7$, $q=1.5$ (same parameters as on Fig. \ref{fig:anisotropy_interpolation}). The blue lines show the cuts along the $y$-axis, where the translations are not broken, while the yellow lines show the cuts along $x$-axis, where the potential is present. A significant suppression of the peak in the direction of translational symmetry breaking is evident.}
		\label{fig:1d_stripes_q_15}
\end{figure}

The result of this interpolation is shown in Fig.\ref{fig:anisotropy_interpolation}. There is no apparent qualitative difference between the modulated and the homogeneous potentials. One can see that both types of the background lead to elliptical stretching of the Fermi surface, and to broadening of the peak in spectral density in the direction of translational symmetry breaking. This broadening is clearly seen on the fixed direction cuts of the spectral density, shown on Fig.\,\ref{fig:1d_stripes_q_15}, where we compare the momentum distribution curves of the spectral density in translationally invariant $y$-direction and translationally broken $x$-direction in both periodic potential and Q-lattice setups.

\section{Discussion}\label{sec:Disc}
Our study shows that the anisotropic broadening of the Fermi surface in holographic models with periodic lattice, pointed out in \cite{Cremonini:2019fzz} and claimed to be relevant for the phenomenology of real exotic condensed matter systems, has the same origin as the damping of quasiparticles due to translational symmetry breaking observed earlier in homogeneous Q-lattices \cite{Ling:2014bda} and Bianchy VII helices \cite{Bagrov:2016cnr}. There are two important consequences which follow from this observation.

First, since the anisotropic decoherence is not related to the periodicity of the lattice, it is insensitive to the structure of the Brillouin zone, and in this way it is essentially a non-Fermi liquid effect. In Fermi liquid, the quasiparticles are only sensitive to the periodic potential via the Umklapp scattering, that is, in the vicinity of the Lifshitz transition; the interaction effects on the Lifshitz transition within the Fermi liquid theory are considered in \cite{katsnelson2000}. The further away the Brillouin zone boundary is, the more Fermi liquid quasiparticles behave just like the single-particle excitations in the vacuum. However, we show that the anisotropic decoherence, on the opposite, gets stronger when the size of the BZ is increased, which 
indicates the non-quasiparticle nature of this phenomenon. 

The holographic models usually give rise to the two kinds of phenomena: one can be derived from the usual behavior of the conventional stable degrees of freedom, propagating in the bulk just like the quasiparticles do on the boundary. The features of this kind are the presence of the Fermi surface, the Umklapp gap, or, in the vector sector, the Drude conductivity peak. There is also another, non-conventional sector, which is due to the effects caused by the black hole horizon deep inside the holographic bulk space. The horizon reflects the deep IR critical nature of the holographic systems \cite{Faulkner:2010tq,Faulkner:2011tm}.  The features of the near-horizon geometry in the bulk govern the unconventional non-Fermi liquid like phenomena in holographic models: these include the unusual power law self-energy of the excitations \cite{Liu:2009dm,reber2015power}, the destruction of the Fermi surface due to translational symmetry breaking \cite{Bagrov:2016cnr} and, for instance, the incoherent conductivity which behaves as a power law of  
temperature \cite{Davison:2015bea,Davison:2014lua}. Our study shows that the anisotropic decoherence is the effect of the second kind, i.e. it is a result of the anisotropic character of the black hole horizon in the models with periodic lattices. 

The analysis of the near-horizon geometries of periodically modulated black holes is extremely hard since it requires using sophisticated numerical techniques. However, we show that the much simpler homogeneous lattice models suit perfectly for the study of the phenomenon under consideration, and this is the second useful consequence of our observation. Only after explicit comparison with the realistic periodic lattice setup we could reliably state that the description provided by the homogeneous lattice is not pathological. 

It is yet an open question which particular features of the holographic horizon play the leading role in the anisotropic destruction of the Fermi surface, and how these features could be interpreted in terms of the quantum critical subpart of the boundary theory. However our study suggests that the first steps in clarifying this may be done in 
relatively simple and accessible models like Q-lattice, where the structure of the near-horizon geometry can be captured in much more detail \cite{Donos:2014uba}, and even 
an analytical treatment is possible in certain cases.   

\section*{Acknowledgements}
We thank Koenraad Schalm, Jan Zaanen, Floris Balm and Aurelio Romero-Bermudez for fruitful discussions on the related topics. 

The work of A.K. is supported by Koenraad Schalm's VICI award of the Netherlands Organization for Scientific Research (NWO), by the Netherlands Organization for Scientific Research/Ministry of Science and Education (NWO/OCW), by the Foundation for Research into Fundamental Matter (FOM). The work of A.B. and M.K. is supported by the Netherlands Organization for Scientific Research (NWO) via the Spinoza Prize of M.K. 

\appendix
\section{\label{app:numerical}The numerical scheme}
In this section we provide some details of the numerical scheme which were not described in Sec. 2 and 3.

The numerical scheme consists of two main parts: calculation of the background metric and the subsequent evaluation of the fermionic Green's function on top of it in the probe limit. Each of these parts, in turn, can be divided into generation of the equations of motions and the boundary conditions, and solving the corresponding boundary problem.

The Einstein and the Dirac equations and their boundary conditions are generated in Mathematica, and the output is parsed into Python 3.7. After that, we use routines described in Sec. 2 and 3.

For the background metric, we use grid of the size $n_x \times n_z = 20\times20$. For such a small grid finite-difference approximation of the derivatives would lead to a very low accuracy, so we use pseudo-spectral method along the radial coordinate $z$ instead. Within this approach, matrices of the derivatives contain information about all grid points. The numerical error of solutions to the Einstein equations is encoded in the trace of the DeTurck vector. In our case, it is of order $10^{-4}$, which is enough for the results to be qualitatively correct.

The fermionic spectral functions are calculated on a grid $n_{k_x}\times n_{k_y}=500\times500$. This gives resolution in the momentum space $\Delta k\approx10^{-2}$, which corresponds to the width of the Fermi surface for fermions with charge $q=1.5$.

The calculations were performed on a laptop with Intel Core i-7 processor and 8 Gb RAM. The recent version of the Python code can be found on \url{https://gitlab.science.ru.nl/iliasov/fermions-on-stripes}.

\section{\label{app:extra_results}Some extra observations}
While we have clearly demonstrated that the anisotropic destruction of the Fermi surface is of the same nature in homogeneous and periodic lattices, it is also interesting to gain a better understanding of why it is the case, and to take a closer look at the differences between the two setting.

\begin{figure}[h]
		\centering
		\includegraphics[width=0.5\textwidth]{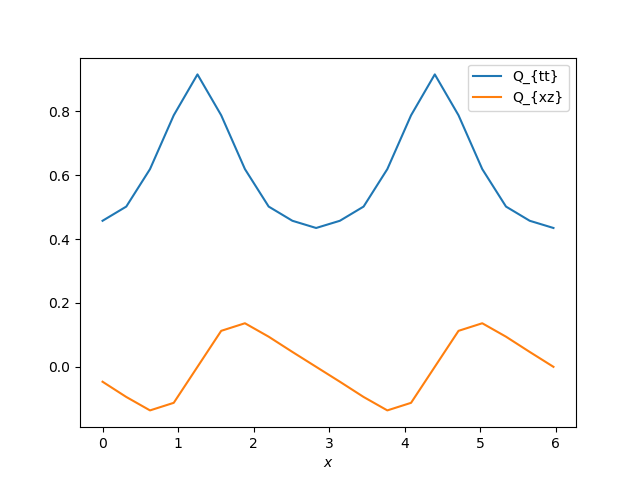}
		\caption{1d slices of $Q_{tt}$ and $Q_{xz}$ components of the metric along the $x$-axis near the conformal boundary, $z=0.16$. The amplitude is taken $V_0=5$, and the wave vector of the lattice is $k_0=0.6$.}
		\label{fig:small_Qxz}
	\end{figure}

The possible reason behind this similarity is the fact that the off-diagonal component of the metric tensor encoded in $Q_{xz}$ function, being exactly zero for the Q-lattice, remains very small for periodic stripes, as one can see in Fig.\ref{fig:small_Qxz}. Therefore the main effect comes from the anisotropic coordinate dependence of the diagonal components of the metric rather than from the explicit modulation.

\begin{figure}[h]
		\includegraphics[width=0.5\textwidth]{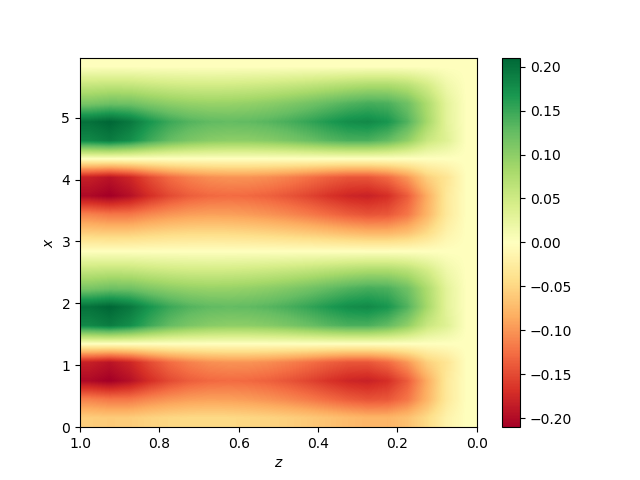}
		\includegraphics[width=0.5\textwidth]{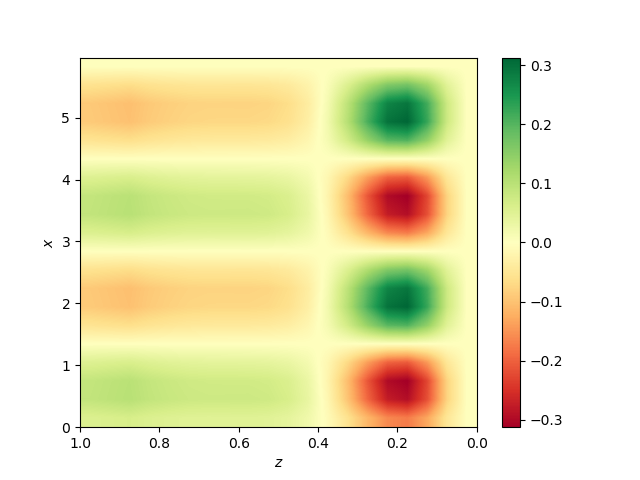}
		
		\caption{Non-diagonal $Q_{xz}$ component of metric for $V_0 =5$, $k_0=0.7$ and $k_0=2.2$.}
		\label{fig:Q_xz}
\end{figure}
However, one can see a mild difference between these cases, -- for the Q-lattice eccentricity of the Fermi surface is a bit larger, while the peak in spectral density is suppressed by the same factor as for the stripe lattice. Unfortunately, we cannot access the regime of very large lattice amplitudes $V_0$ because of arising numerical instabilities. Still, we can speculate that a larger off-diagonal component of the metric tensor specific for the striped background would lead to reducing the effect of geometrical stretching while retaining the anisotropic decoherence.

Interesting to note that, although the transition from a circular Fermi surface to a stretched and smeared one is governed by diagonal components of the metric, a clear signature of this transition is evident in the off-diagonal component $Q_{xz}$. In Fig.\ref{fig:Q_xz}, one can see that when $k_0$ is driven to higher values, profile of $Q_{xz}$ function undergoes a change. 

\section{\label{app:q1}Spectral function at smaller charge of the fermion}
\begin{figure}[p]
		\includegraphics[width=0.5\textwidth]{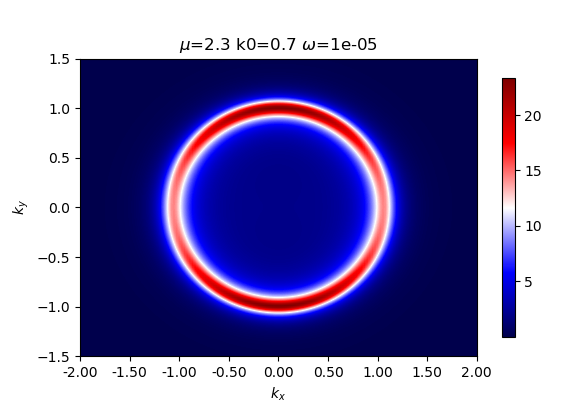}
		\includegraphics[width=0.5\textwidth]{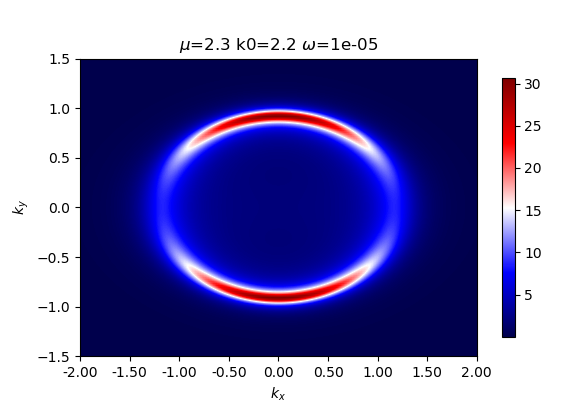}
		\caption{Spectral density for the Q-lattice, $\theta=\pi/4$, at $k_0=0.7$ and $k_0=2.2$, $V_0=7$, $q=1$.}
		\label{fig:anisotropy_1}
\end{figure}

\begin{figure}[p]
		\includegraphics[width=0.5\textwidth]{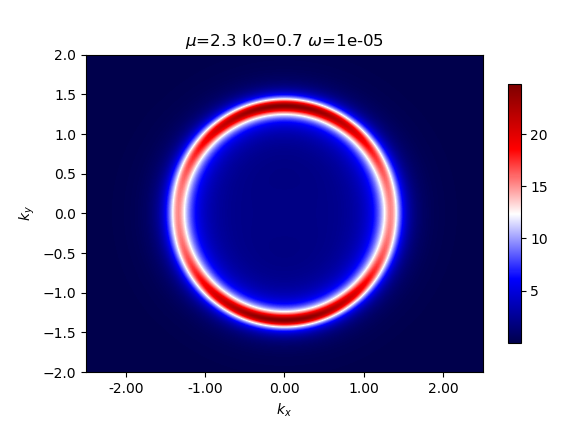}
		\includegraphics[width=0.5\textwidth]{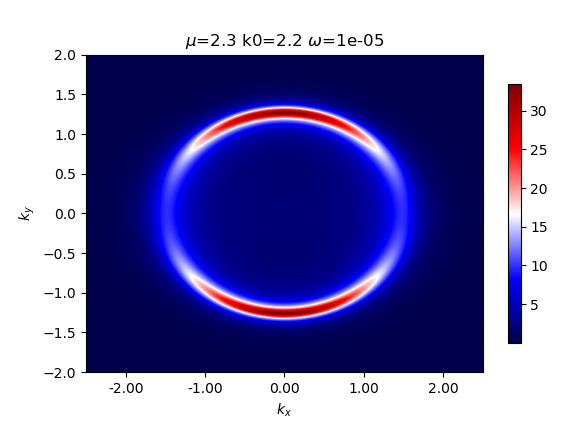}
		\caption{Spectral function for the interpolating metric, $\theta=\pi/8$, at $k_0=0.7$ (left) and $k_0=2.2$ (right), $V_0=7$, $q=1$.}
		\label{fig:anisotropy_05}
\end{figure}

\begin{figure}[p]
		\includegraphics[width=0.5\textwidth]{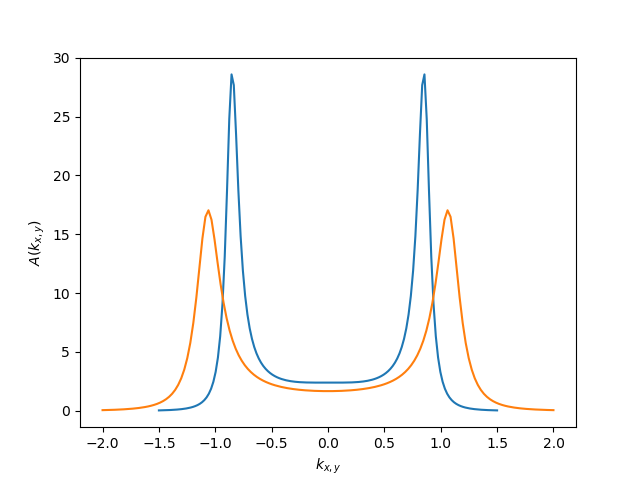}
				\includegraphics[width=0.5\textwidth]{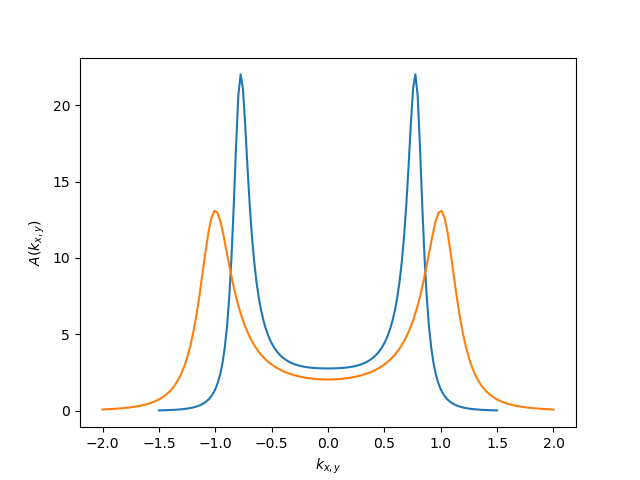}
		\caption{1d sections of the spectral function for stripes (left) and for the Q-lattice (right) at $k_0=2.2$, $V_0=7$, $q=1$.}
		\label{fig:1d_stripes_q10}
\end{figure}

\begin{figure}[p]
		\centering
		\includegraphics[width=0.9\textwidth]{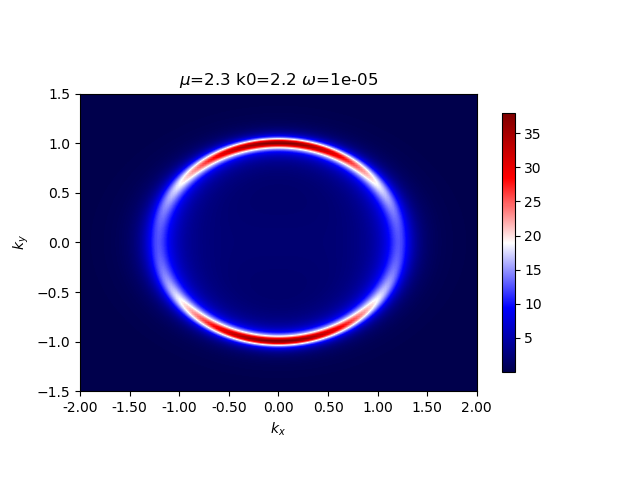}
		\caption{Spectral density for the striped background, $\theta=0$, at $k_0=2.2$ and $q=1$.}
		\label{fig:Stripes_q1_k02_2}
	\end{figure}

Finally, it is instructive to look at how the Fermi surface looks at smaller electric charge of the fermion, $q=1$. In this case, the quasiparticle peak is smeared, and the effect of anisotropic destruction is weaker, but on the other hand it gives a better visualization of the aforediscussed transitions. We provide a series of plots (Figs. \ref{fig:anisotropy_1} -- \ref{fig:Stripes_q1_k02_2}) without any additional comments.

\providecommand{\href}[2]{#2}\begingroup\raggedright

\endgroup


\clearpage

\begin{thebibliography}{40}

\bibitem{Zaanen:2015oix}
J.~Zaanen, Y.-W. Sun, Y.~Liu, and K.~Schalm, {\em {Holographic Duality in
  Condensed Matter Physics}},
\newblock Cambridge Univ. Press, 2015.

\bibitem{keimer2015quantum}
B.~Keimer, S.~A. Kivelson, M.~R. Norman, S.~Uchida and J.~Zaanen, {\it From
  quantum matter to high-temperature superconductivity in copper oxides},  {\em
  Nature} {\bf 518} (2015), no.~7538 179.

\bibitem{Cubrovic:2009ye}
M.~Cubrovic, J.~Zaanen and K.~Schalm, {\it {String Theory, Quantum Phase
  Transitions and the Emergent Fermi-Liquid}},  {\em Science} {\bf 325} (2009)
  439--444 [\href{http://arXiv.org/abs/0904.1993}{{\tt 0904.1993}}].

\bibitem{Cubrovic:2011xm}
M.~Cubrovic, Y.~Liu, K.~Schalm, Y.-W. Sun and J.~Zaanen, {\it {Spectral probes
  of the holographic Fermi groundstate: dialing between the electron star and
  AdS Dirac hair}},  {\em Phys. Rev.} {\bf D84} (2011) 086002
  [\href{http://arXiv.org/abs/1106.1798}{{\tt 1106.1798}}].

\bibitem{Liu:2009dm}
H.~Liu, J.~McGreevy and D.~Vegh, {\it {Non-Fermi liquids from holography}},
  {\em Phys. Rev.} {\bf D83} (2011) 065029
  [\href{http://arXiv.org/abs/0903.2477}{{\tt 0903.2477}}].

\bibitem{Faulkner:2009wj}
T.~Faulkner, H.~Liu, J.~McGreevy and D.~Vegh, {\it {Emergent quantum
  criticality, Fermi surfaces, and AdS(2)}},  {\em Phys. Rev.} {\bf D83} (2011)
  125002 [\href{http://arXiv.org/abs/0907.2694}{{\tt 0907.2694}}].

\bibitem{Faulkner:2010tq}
T.~Faulkner and J.~Polchinski, {\it {Semi-Holographic Fermi Liquids}},  {\em
  JHEP} {\bf 06} (2011) 012 [\href{http://arXiv.org/abs/1001.5049}{{\tt
  1001.5049}}].

\bibitem{Faulkner:2011tm}
T.~Faulkner, N.~Iqbal, H.~Liu, J.~McGreevy and D.~Vegh, {\it {Holographic
  non-Fermi liquid fixed points}},  {\em Phil. Trans. Roy. Soc.} {\bf A 369}
  (2011) 1640 [\href{http://arXiv.org/abs/1101.0597}{{\tt 1101.0597}}].

\bibitem{fu2006dichotomy}
H.~Fu and D.-H. Lee, {\it Dichotomy between the nodal and antinodal excitations
  in high-temperature superconductors},  {\em Physical Review B} {\bf 74}
  (2006), no.~17 174513.

\bibitem{kanigel2006evolution}
A.~Kanigel, M.~Norman, M.~Randeria, U.~Chatterjee, S.~Souma, A.~Kaminski,
  H.~Fretwell, S.~Rosenkranz, M.~Shi, T.~Sato {\em et.~al.}, {\it Evolution of
  the pseudogap from fermi arcs to the nodal liquid},  {\em Nature Physics}
  {\bf 2} (2006), no.~7 447.
  
\bibitem{katsnelson2008}  
M. I. Katsnelson, V. Yu. Irkhin, L. Chioncel, A. I. Lichtenstein, and R. A. de Groot,
{\it Half-metallic ferromagnets: From band structure to many-body effects}, {\em Rev. Mod. Phys.}
{\bf 80} (2008), no.~2 315. 

\bibitem{Horowitz:2012ky}
G.~T. Horowitz, J.~E. Santos and D.~Tong, {\it {Optical Conductivity with
  Holographic Lattices}},  {\em JHEP} {\bf 07} (2012) 168
  [\href{http://arXiv.org/abs/1204.0519}{{\tt 1204.0519}}].

\bibitem{Horowitz:2012gs}
G.~T. Horowitz, J.~E. Santos and D.~Tong, {\it {Further Evidence for
  Lattice-Induced Scaling}},  {\em JHEP} {\bf 11} (2012) 102
  [\href{http://arXiv.org/abs/1209.1098}{{\tt 1209.1098}}].

\bibitem{Donos:2014yya}
A.~Donos and J.~P. Gauntlett, {\it {The thermoelectric properties of
  inhomogeneous holographic lattices}},  {\em JHEP} {\bf 01} (2015) 035
  [\href{http://arXiv.org/abs/1409.6875}{{\tt 1409.6875}}].

\bibitem{Rangamani:2015hka}
M.~Rangamani, M.~Rozali and D.~Smyth, {\it Spatial modulation and
  conductivities in effective holographic theories},  {\em JHEP} {\bf 07}
  (2015) 024 [\href{http://arXiv.org/abs/1505.05171}{{\tt 1505.05171}}].

\bibitem{Liu:2012tr}
Y.~Liu, K.~Schalm, Y.-W. Sun and J.~Zaanen, {\it {Lattice Potentials and
  Fermions in Holographic non Fermi-Liquids: Hybridizing Local Quantum
  Criticality}},  {\em JHEP} {\bf 10} (2012) 036
  [\href{http://arXiv.org/abs/1205.5227}{{\tt 1205.5227}}].

\bibitem{Ling:2013aya}
Y.~Ling, C.~Niu, J.-P. Wu, Z.-Y. Xian and H.-B. Zhang, {\it {Holographic
  Fermionic Liquid with Lattices}},  {\em JHEP} {\bf 07} (2013) 045
  [\href{http://arXiv.org/abs/1304.2128}{{\tt 1304.2128}}].

\bibitem{cremonini2018holographic}
S.~Cremonini, L.~Li and J.~Ren, {\it Holographic fermions in striped phases},
  {\em Journal of High Energy Physics} {\bf 2018} (2018), no.~12 80.

\bibitem{Cremonini:2019fzz}
S.~Cremonini, L.~Li and J.~Ren, {\it {Spectral Weight Suppression and Fermi
  Arc-like Features with Strong Holographic Lattices}},
  \href{http://arXiv.org/abs/1906.02753}{{\tt 1906.02753}}.

\bibitem{Andrade:2013gsa}
T.~Andrade and B.~Withers, {\it A simple holographic model of momentum
  relaxation},  {\em JHEP} {\bf 05} (2014) 101
  [\href{http://arXiv.org/abs/1311.5157}{{\tt 1311.5157}}].

\bibitem{Donos:2013eha}
A.~Donos and J.~P. Gauntlett, {\it Holographic q-lattices},  {\em JHEP} {\bf
  04} (2014) 040 [\href{http://arXiv.org/abs/1311.3292}{{\tt 1311.3292}}].

\bibitem{Donos:2012js}
A.~Donos and S.~A. Hartnoll, {\it {Interaction-driven localization in
  holography}},  {\em Nature Phys.} {\bf 9} (2013) 649--655
  [\href{http://arXiv.org/abs/1212.2998}{{\tt 1212.2998}}].

\bibitem{Ling:2014bda}
Y.~Ling, P.~Liu, C.~Niu, J.-P. Wu and Z.-Y. Xian, {\it {Holographic fermionic
  system with dipole coupling on Q-lattice}},  {\em JHEP} {\bf 12} (2014) 149
  [\href{http://arXiv.org/abs/1410.7323}{{\tt 1410.7323}}].

\bibitem{Bagrov:2016cnr}
A.~Bagrov, N.~Kaplis, A.~Krikun, K.~Schalm and J.~Zaanen, {\it {Holographic
  fermions at strong translational symmetry breaking: a Bianchi-VII case
  study}},  {\em JHEP} {\bf 11} (2016) 057
  [\href{http://arXiv.org/abs/1608.03738}{{\tt 1608.03738}}].

\bibitem{geim2013}
A. K. Geim and I. V. Grigorieva, {\it Van der Waals heterostructures}, {\em Nature} {\bf 499} (2013), 419.

\bibitem{dean2010}
C. R. Dean,  A. F. Young,  I. Meric, C. Lee, L. Wang, S. Sorgenfrei, K. Watanabe, T. Taniguchi, P. Kim, K. L. Shepard,  and J. Hone, {\it Boron nitride substrates for high-quality graphene electronics},  {\em Nature Nanotech.} {\bf 5} (2010), 722.

\bibitem{xue2011} 
J. Xue, J. Sanchez-Yamagishi, D. Bulmash, P. Jacquod, A. Deshpande, K. Watanabe, T. Taniguchi, P. Jarillo-Herrero, and  B. J. LeRoy, {\it Scanning tunnelling microscopy and spectroscopy of ultra-flat graphene on hexagonal boron nitride}, {\em Nature Mater.}  {\bf 10} (2011), 282.

\bibitem{woods2014}
C. R. Woods, L. Britnell, A. Eckmann, R. S. Ma, J. C. Lu, H. M. Guo, X. Lin, G. L. Yu, Y. Cao, R. V. Gorbachev, A. V. Kretinin,  J. Park, L. A. Ponomarenko, M. I. Katsnelson, Yu. N. Gornostyrev, K. Watanabe, T. Taniguchi, C. Casiraghi, H.-J. Gao, A. K. Geim,  and K. S. Novoselov, {\it Commensurate-incommensurate transition in graphene on hexagonal boron nitride} {\em Nature Phys.} {\bf 10} (2014), 451.  

\bibitem{cao1}
Y. Cao, V. Fatemi, A. Demir, S. Fang, S. L. Tomarken, J. Y. Luo, J. D. Sanchez-Yamagishi, K. Watanabe, T. Taniguchi, E. Kaxiras, R. C. Ashoori,  and P. Jarillo-Herrero, {\it Correlated insulator behaviour at half-filling in magic-angle graphene superlattices} {\em Nature} {\bf 556} (2018), 80. 

\bibitem{cao2}
Y. Cao, V. Fatemi, S. Fang, K. Watanabe, T. Taniguchi, E. Kaxiras, and P. Jarillo-Herrero, {\it Unconventional superconductivity in magic-angle graphene superlattices} {\em Nature} {\bf 556} (2018), 43. 

\bibitem{cao3}
Y. Cao, D. Chowdhury, D. Rodan-Legrain, O. Rubies-Bigord\'a, K. Watanabe, T. Taniguchi, T. Senthil, and P. Jarillo-Herrero, {\it Strange metal in magic-angle graphene with near Planckian dissipation}, {\em arXiv:1811.04920}.

\bibitem{wallbank2013}
J. R. Wallbank, A. A. Patel, M. Mucha-Kruczy\'nski, A. K. Geim, and V. I. Fal'ko,  {\it Generic miniband structure of graphene on a hexagonal substrate},  {\em Phys. Rev. B} 87 (2013), 245408.

\bibitem{slotman2015}
G. J. Slotman, M. M. van Wijk, P.-L. Zhao, A. Fasolino, M. I. Katsnelson, and S. Yuan, {\it Effect of structural relaxation on the electronic structure of graphene on hexagonal boron nitride} {\em Phys. Rev. Lett.} {\bf 115} (2015), 186801. 

\bibitem{vanacore2014minding}
G.~Vanacore and P.~W. Phillips, {\it Minding the gap in holographic models of
  interacting fermions},  {\em Physical Review D} {\bf 90} (2014), no.~4
  044022.

\bibitem{BalmLattice}
 F. Balm, A. Krikun, A. Romero-Berm\'udez, K. Schalm and J. Zaanen, {\it Isolated zeros destroy Fermi surface in holographic models with a lattice}, 
\href{http://arXiv.org/abs/1909.09394}{{\tt 1909.09394}}.

\bibitem{Headrick:2009pv}
M.~Headrick, S.~Kitchen and T.~Wiseman, {\it A new approach to static numerical
  relativity, and its application to kaluza-klein black holes},  {\em Class.
  Quant. Grav.} {\bf 27} (2010) 035002
  [\href{http://arXiv.org/abs/0905.1822}{{\tt 0905.1822}}].

\bibitem{Wiseman:2011by}
T.~Wiseman, {\em Numerical construction of static and stationary black holes}.
\newblock 2011.

\bibitem{Adam:2011dn}
A.~Adam, S.~Kitchen and T.~Wiseman, {\it A numerical approach to finding
  general stationary vacuum black holes},  {\em Class. Quant. Grav.} {\bf 29}
  (2012) 165002 [\href{http://arXiv.org/abs/1105.6347}{{\tt 1105.6347}}].

\bibitem{Son:2002sd}
D.~T. Son and A.~O. Starinets, {\it {Minkowski space correlators in AdS / CFT
  correspondence: Recipe and applications}},  {\em JHEP} {\bf 09} (2002) 042
  [\href{http://arXiv.org/abs/hep-th/0205051}{{\tt hep-th/0205051}}].
\bibitem{lifshitz1973}
I. M. Lifshitz, M. Ya. Azbel, and M. I. Kaganov, {\it Electron Theory of Metals} (Consultants Bureau, New York, 1973).

\bibitem{vonsovsky1989}
S. V. Vonsovsky and M. I. Katsnelson, {\it Quantum Solid-State Physics} (Springer, New York, 1989).

\bibitem{katsnelson2000}
M. I. Katsnelson and A. V. Trefilov, {\it Fermi-liquid theory of electronic topological transitions and screening anomalies in metals} {\em Phys. Rev. B} {\bf 61} (2000), 1643.

\bibitem{reber2015power}
T.~Reber, X.~Zhou, N.~Plumb, S.~Parham, J.~Waugh, Y.~Cao, Z.~Sun, H.~Li,
  Q.~Wang, J.~Wen {\em et.~al.}, {\it Power law liquid-a unified form of
  low-energy nodal electronic interactions in hole doped cuprate
  superconductors},  
   [\href{http://arXiv.org/abs/1509.01611}{{\tt 1509.01611}}].

\bibitem{Davison:2015bea}
R.~A. Davison and B.~Gout\'eraux, {\it {Dissecting holographic conductivities}},
   {\em JHEP} {\bf 09} (2015) 090 [\href{http://arXiv.org/abs/1505.05092}{{\tt
  1505.05092}}].

\bibitem{Davison:2014lua}
R.~A. Davison and B.~Gout\'eraux, {\it {Momentum dissipation and effective
  theories of coherent and incoherent transport}},  {\em JHEP} {\bf 01} (2015)
  039 [\href{http://arXiv.org/abs/1411.1062}{{\tt 1411.1062}}].

\bibitem{Donos:2014uba}
A.~Donos and J.~P. Gauntlett, {\it Novel metals and insulators from
  holography},  {\em JHEP} {\bf 06} (2014) 007
  [\href{http://arXiv.org/abs/1401.5077}{{\tt 1401.5077}}].


\end{thebibliography}
\end{document}